\documentstyle[prb,aps,preprint,tighten]{revtex}
\begin{document}
\draft

\title{
Local Disorder In the Oxygen Environment Around Praseodymium in \\
Y$_{1-x}$Pr$_{x}$Ba$_{2}$Cu$_{3}$O$_{7}$ from X-Ray Absorption Fine
Structure
}
\author{C.  H.  Booth, F.  Bridges}
\address{
Physics Department, 
University of California,
Santa Cruz, CA 95064
}
\author{J. B. Boyce}
\address{
Xerox Palo Alto Research Center, Palo Alto, CA 94304
}
\author{T. Claeson}
\address{
Physics Department,
Chalmers Univ.  of Tech.,
S-41296 Gothenburg, Sweden 
}
\author{Z. X. Zhao}
\address{
National Laboratory for Superconductivity,
Institute of Physics,
Academia Sinica,
Beijing 100080, China 
}
\author{Phillip Cervantes}
\address{
Physics Department,
University of California,
Santa Cruz, CA 95064
}
\date{\today}

\maketitle
\begin{abstract}
PrBa$_{2}$Cu$_{3}$O$_{7}$ (PBCO) is unique in the
RBa$_{2}$Cu$_{3}$O$_{7}$ (R=rare earth) series because it is not
superconducting. In fact, for
Y$_{1-x}$Pr$_{x}$Ba$_{2}$Cu$_{3}$O$_{7}$, T$_{c}$ drops monotonically
with Pr concentration, with T$_{c}$ going to zero at $\sim$55\% Pr.
There  have been many studies of this material with the hope that an
explanation for the lack of superconductivity in PBCO might help
explain why YBa$_{2}$Cu$_{3}$O$_{7}$ (YBCO) is superconducting.
Explanations  center around hole localization, requiring an extra hole
on the Y (Pr) site or a localized hole in the O 2{\it p} shell, at the
expense of a mobile hole in the Cu-O planes.  To help provide clues
that could point to a particular model, and to search for anomalies in
the local structure, we present K-edge XAFS (X-Ray Absorption Fine
Structure) data for various concentrations of Pr in Y$_{1-
x}$Pr$_{x}$Ba$_{2}$Cu$_{3}$O$_{7}$. The character of the Pr K-edge
XAFS data indicate that most of the Pr substitutes onto the Y site and
is well ordered with respect to the unit cell.  These data also show
that the amplitude of the first Pr-O peak is greatly reduced when
compared to the first Y-O peak in pure YBCO, and decreases with
increasing Pr concentration.  In contrast, the Y K-edge data for these
alloys show little if any change in the oxygen environment, while the
Cu K-edge data show a 10\% reduction in the first Cu-O peak.  Fits to
the Pr data suggest that some oxygen atoms about the Pr become
disordered and/or distorted; most of the Pr-O nearest-neighbor
distances are 2.45  \AA, but about 15-40\% of them are in a
possibly broadened peak at 2.27$^{+0.03}_{-0.12}$  \AA.  The Cu K-edge
XAFS data show a slight broadening but no loss of oxygens, which is
consistent with a radial distortion of the Pr-O bond.  The existence
and the size of these two bond lengths is consistent with a mixture of
Pr$^{3+}$ and Pr$^{4+}$ bonds, and to a formal valence of
+3.33$^{+0.07}_{-0.18}$ for the Pr ion.
\end{abstract}
\narrowtext

\section{Introduction}
\label{introduction}
An interesting property of YBa$_{2}$Cu$_{3}$O$_{7}$ (YBCO) is that one
may replace yttrium with almost any rare earth and still obtain a
high-T$_{c}$ superconductor,\cite{Hor87,Schneemeyer87} isostructural
with YBCO.  However, there are two rare earths that do not form a
high-T$_{c}$ superconductor when attempts are made to substitute them
onto the Y site: cerium and praseodymium.\cite{Murphy87,Soderholm87}
Attempts  to substitute with Ce have not produced the YBCO structure,
even in small concentrations.  Praseodymium, however, substitutes
readily onto the Y site, with T$_{c}$ dropping monotonically with Pr
concentration. Therefore, PrBa$_{2}$Cu$_{3}$O$_{7}$ (PBCO) is a unique
material in that it has the YBCO structure, with a rare earth instead
of Y, yet is not superconducting.  In contrast, substitution by Gd,
for example, 
has essentially no change in T$_{c}$ or in the crystal
structure.\cite{Boyce87} There has been a large effort to explain this
reduction in T$_{c}$ as a new avenue in the search to explain why YBCO
is superconducting in the first place.  PBCO is interesting in its own
right, however, as an insulating layer between layers of YBCO in
multilayer films, or as a well matched substrate for epitaxial
films of YBCO.

There are many related explanations for the decrease in T$_{c}$ in
Y$_{1-x}$Pr$_{x}$Ba$_{2}$Cu$_{3}$O$_{7}$ (YBCO:Pr),
mostly involving the localization of holes through valence change or
magnetic pair breaking mechanisms.  The earliest arguments note that
Ce, Pr and Tb are the only rare earths that can exist in either a +3
or +4 valent state.\cite{Soderholm87}  Since Y only exists in a +3
valent state, one explanation for the decrease in T$_{c}$ with Pr
concentration is that some fraction of the Pr exists in its formal +4
state, filling a hole state in the Cu-O planes and localizing it onto
the Pr site.  There is some experimental evidence to support this
view, notably the measured magnetic susceptibility of 2.7  $\mu_{B}$
per Pr ion in PBCO,\cite{Dalichaouch88} which may be compared to the
Pr$^{3+}$ ion (3.58  $\mu_{B}$) and the Pr$^{4+}$ ion (2.54
$\mu_{B}$).  However, this result should be considered in light of
some model systems that give moments significantly different than the
ionic moments, such as the moment of BaPrO$_{3}$ (0.7
$\mu_{B}$),\cite{Bickel88}  where Pr$^{4+}$ is presumably the natural
state.

Another possible mechanism for suppressing T$_{c}$ in this material
utilizes the fact that, except for Ce, the Pr$^{3+}$ ion is the
largest of all the rare earth +3 ions.  The 4{\it f}$^{2}$ electrons
are far enough extended to hybridize with the O 2{\it p} electrons,
and thus interfere with the conduction band.\cite{Soderholm91}  In
fact, crystal field measurements can be reconciled with Pr in a
completely +3 state,\cite{Soderholm91,Jostarndt92} with broad
transitions that can be attributed to hybridization between the Cu
5{\it d}-O 2{\it p} states and the Pr 4{\it f}
states.\cite{Jostarndt92}

X-Ray Absorption Near-Edge Structure studies (XANES) also indicate
that Pr is primarily trivalent.  Comparisons of the Pr L$_{III}$-edge
of PBCO to that for PrO$_{2}$ and Pr$_{2}$O$_{3}$ indicate a valence
close to +3,\cite{Alp88} or slightly above it.\cite{Horn87,Lytle90}
The  latter studies suggest that a small amount of Pr +4 is present;
Horn {\it et al.}  \cite{Horn87} give a mean valence of +3.1,  while
Lytle {\it et al.}  \cite{Lytle90} find that the valence decreases
from +3.45  for YBCO:Pr 20\% to +3.25  for YBCO:Pr 60\% with all of
the Pr$^{4+}$ on the Cu(2) site (see Fig. \ref{fig1}  for structural notation).
M-edge  studies \cite{Neukirch88} and electron-energy loss
spectroscopy (EELS) studies \cite{Fink88} also find that the Pr
valence is close to +3.  These same studies show that the number of
holes in the CuO$_{2}$ planes does not change with Pr concentration,
while normal-state Hall effect measurements \cite{Jia92} show a
decrease in the number of mobile charge carriers with x, indicating
that the holes are being localized rather than filled.  Photoemission
experiments also find that the Pr valence is close to +3 and that
there is significant hybridization of the Pr 4{\it f} states with
other valence states.\cite{Kang89}  In addition, thin films of
Y$_{0.5}$Tb$_{0.5}$Ba$_{2}$Cu$_{3}$O$_{7}$ have
recently been produced, with a T$_{c}$ of 92 K.\cite{Cao93}  Such a
high T$_{c}$ for a 50\% terbium sample brings into question the
argument that the suppression of T$_{c}$ arises solely because two
valence states can exist.

A theory which combines many of these ideas, proposed by Fehrenbacher
and Rice,\cite{Fehrenbacher93} shows that three electronic
configurations are favored: a Pr$^{3+}$(4{\it f}$^{2}$)-O state, a
Pr$^{4+}$(4{\it f}$^{1}$)-O state, and a Pr$^{3+}$(4{\it
f}$^{2}$$\underline{L}$)-O, where the Pr$^{3+}$(4{\it
f}$^{2}\underline{L}$) indicates that the hole does not reside within
the praseodymium's electronic structure; rather, the hole is in the
ligand. The low-energy ligand hole is obtained via a superposition of
O 2{\it p}$_{\pi}$ orbitals with {\it f} symmetry.  The most
energetically favorable configuration has Pr$^{3+}$ 60-70\% of the
time, with the remaining Pr-O bonds split between the other two
configurations. They reason that such a mixture would be consistent
with all the above measurements, since ionic Pr$^{4+}$ would only
exist 15-20\% of the time, even though formally the Pr is in a +4
state 30-40\% of the time.

There is also some structural evidence from neutron and x-ray
diffraction for a mixed Pr valency.  Diffraction results show that
PBCO is isostructural with YBCO.  The diffraction result for the
average Pr-O nearest-neighbor distance of 2.45  {\AA}
\cite{Morales90,Neumeier90,Lowe-Ma92,Guillaume93} is
reasonably consistent with Pr$^{3+}$-O bond lengths in other Pr
oxides, such as in bixbyite Pr$_{2}$O$_{3}$, which has a range of Pr-O
bond lengths from 2.33  {\AA} to 2.66  \AA.\cite{Wyckoff}  It has been
noted, however, that both the Cu-O plane separation
\cite{Neumeier90,Guillaume93}
and the Pr-O(2) and Cu(2)-O(3) bond lengths\cite{Guillaume93} are more
consistent with a
valence in the 
range of +3.3-3.4,  as determined by interpolating trends in these
bond lengths with rare-earth ionic radius.  However, this
interpolation argument has at least two possible flaws.  First, just
because the trend with ionic radius is no longer linear for Pr does
not necessarily mean Pr is in a mixed valence state.  The Pr$^{3+}$
ion is the largest ion anyone has been able to place between the Cu-O
planes in this structure, so new forces, including more oxygens
bonding to the Pr than, say, Y, may be coming into play.  In addition,
there are several compounds that show a wide range of Pr$^{3+}$ bond
lengths, such as bixbyite Pr$_{2}$O$_{3}$.  Secondly, if three Pr-O
states exist, as suggested by Fehrenbacher and
Rice,\cite{Fehrenbacher93} then one would expect to see two, and
perhaps three, different Pr-O bond lengths; Pr$^{4+}$-O bonds range in
length from 2.18  {\AA} in BaPrO$_{3}$ to 2.32  {\AA} in PrO$_{2}$.
Therefore  the three Pr-O states should have bond lengths made up of a
long Pr$^{3+}$(4{\it f}$^{2}$)-O, a short Pr$^{4+}$(4{\it f}$^{1}$)-O,
and a Pr$^{3+}$(4{\it f}$^{2}${\underline L})-O (formally,
Pr$^{4+}$-O), whose bond length
should be closer
to Pr$^{4+}$(4{\it f}$^{1}$)-O.  Diffraction would not be sensitive to
such a distribution.  It will either see the dominant bond length
(perhaps the Pr$^{3+}$(4{\it f}$^{2}$)-O), or an average bond length
consistent with a mixed valency, depending on the degree of disorder.

Likewise, trends in the local O environment around the Pr with Pr
concentration are only obtainable with large Pr concentrations.  If
trends exist in the concentration region where YBCO:Pr is still
superconducting, that is, Pr concentrations $\lesssim$ 55\%,
diffraction may not see them.  Therefore diffraction results do not
rule out disorder and/or distortions of the O around the Pr.  A
distortion in the Pr local structure could indicate the presence of
some formal or ionic Pr$^{4+}$, but other disorder may also exist.
For  instance, there is experimental evidence from x-ray diffraction
to show that the Cu(2)-O(4)
distance is anomalously 
short,\cite{Lowe-Ma92} and therefore may impede charge transfer
between the planes and the chains.  The length of this bond is
consistent with the trend with ionic radius of
Pr$^{3+}$,\cite{Guillaume93} but it is still the shortest Cu(2)-O(4)
bond measured in the YBCO-type materials.  Any of these results could
produce some localization of charge carriers.

Since local distortions and/or disorder in the oxygen environment
around praseodymium may indicate the presence of some Pr with a formal
valence of 4+, we have conducted X-Ray Absorption Fine Structure
(XAFS) experiments on several concentrations of
Y$_{1-x}$Pr$_{x}$Ba$_{2}$Cu$_{3}$O$_{7}$, at the Cu, Y, and Pr K-edges
(E$_{o} \simeq$ 9 keV, 17 keV, and 42 keV, respectively).  (A
preliminary version of this work has recently appeared in the
literature.\cite{Booth93})
XAFS is ideally suited to this problem, because it can give us precise
information about the local structure around the absorbing atom,
including average bond distances and harmonic broadening factors.  In
particular, we should be able to determine exactly which site the Pr
is occupying and observe changes in the XAFS compared to YBCO.  By
taking data at three different absorption edges, we can further
constrain our results for the CuO$_{2}$ planes and get more
information regarding the O(4) site.

In Sec.  \ref{data} we describe some details of the sample preparation
and data collection.  We explain our data reduction procedures and
describe some visible features of the data in Sec.  \ref{features}.
Our  analysis and fit results are detailed in Sec.  \ref{fits}.  We
discuss the implications of our results in Sec.  \ref{discussion} and
give a brief conclusion of our findings in Sec.  \ref{conclusion}.

\section{Samples and data collection}
\label{data}

\subsection{\mbox{XAFS} Samples}
All the samples were prepared by the same procedure as described in
Ref. \onlinecite{Zhu92}.  Stoichiometric amounts of CuO, BaCO$_{3}$,
Y$_{2}$O$_{3}$, and Pr$_{6}$O$_{11}$ were mixed, ground, and heated at
920$^{o}$C in air for 20 hours.  They were then reground, pressed into
pellets, calcined at 940$^{o}$C in an oxygen atmosphere, and slowly
cooled at 60$^{o}$C/h to room temperature.
All the  Y$_{1-x}$Pr$_{x}$Ba$_{2}$Cu$_{3}$O$_{7-y}$ 
samples were characterized by x-ray diffraction using Cu K$\alpha$
radiation. The patterns indicate that all the samples with
x $\leq$ 
0.6  are essentially single phase.  The 100\% Pr sample may have small
amounts of phase impurities, but they are too small to be seen in the
XAFS (see Sec.  \ref{fits}).  Oxygen content was analyzed by an
improved gaseous volumetric technique and found to be within the range
of 6.95$\pm$0.02  for all samples.  In all the compounds T$_{c}$ was
determined by a standard four-probe method.  Superconducting phase
purity was checked using an AC mutual inductance bridge.  Only one
step was observed for all superconducting samples.  Fig.  \ref{fig1}
shows T$_{c}$ as a function of Pr concentration and Table \ref{table1}
gives the lattice parameters and the oxygen content measurements for
these samples.

XAFS scans were performed on the 0, 30, 50, and 100\% Pr samples at
about 80 K for the Pr K-edge and about 50 K for the Cu and Y K-edges.
To  prepare the samples for these transmission experiments, we
reground the pellets and sifted the powder through a 30 $\mu$m filter
onto pieces of scotch tape.  The tape was then cut and stacked into
multiple layers to obtain samples of thickness, t, such that $\mu$t $
\simeq 1$ at the absorption edge, where $\mu$ is the absorption
coefficient. All XAFS experiments were performed at SSRL on beamlines
7-3 and 10-2.  Measurements of the incident and transmitted
intensities were made with {gas-ionization chambers}.  Si (111)
monochromator crystals were used for the Cu and Y K-edge experiments,
while Si (400) crystals were required for the Pr
K-edge.

\section{General Features of the data}
\label{features}
\subsection{Standard reduction and comparison of different absorption
edges} Y, Pr, and Cu K-edge data for $\chi$(k) = ($\mu$ -
$\mu_{o}$)/$\mu_{o}$
show XAFS
oscillations beyond 16 \AA$^{-1}$ for all K-edges.  Fig.  \ref{fig2}
shows k$\chi$(k) for all three K-edges for
Y$_{0.5}$Pr$_{0.5}$Ba$_{2}$Cu$_{3}$O$_{7}$. The ``free atom
absorption'' $\mu_o$ is
determined in the usual way \cite{Hayes82} by fitting a spline
function or a polynomial through the data above the absorption edge.
The  photoelectron wave vector k is given by k=
$\sqrt{(2m_{e}/\hbar^2)(E-E_{o})}$, where E$_{o}$ is the K-edge
threshold energy.  This procedure was modified somewhat for the Pr
K-edge data, as detailed in part B of this section.

The most striking difference between the Y and the Pr K-edge XAFS in
Fig. \ref{fig2} is the overall drop in amplitude of the Pr K-edge
XAFS, particularly in the low k part of the spectrum.  The amplitudes
of the high-k XAFS are similar.  The Cu K-edge XAFS are comparable to
data we have previously obtained on YBCO.

The Fourier transform (FT) of the data yields peaks in r-space
corresponding to different radial distances from the excited atom.
The  peaks in the FT
of k$\chi$(k)
are shifted in distance due to phase shifts at the scattering and
absorbing atoms, and must be compared to standard compound XAFS to
obtain the bond lengths.  In the figures showing FT data, the shifted
position of the peaks are given by vertical lines.  In Fig.
\ref{fig3}a  and \ref{fig3}b, the FT of k$\chi$(k) for data collected
on PBCO at the Pr K-edge and for CeO$_{2}$ at the Ce K-edge are
compared. FT's for several concentrations of Pr in YBCO:Pr for the Y,
Pr, and Cu K-edges are presented in Fig.  \ref{fig4}, \ref{fig5},
and \ref{fig6}.

By comparing the FT of the XAFS from the various edges, we can quickly
ascertain the dominant substitution site for Pr.  From the solid line
in Fig.  \ref{fig3}a (PBCO, Pr K-edge) and the solid line in Fig.
\ref{fig4}
(YBCO, Y
K-edge) we can see by the location of the first two peaks that the
oxygen (first peak) and copper (second peak) environment around Pr in
PBCO is very similar to that around Y in YBCO.  In contrast, a
simulation using FEFF5 \cite{FEFF5} of the Ba K-edge XAFS in YBCO puts
the oxygen and copper peaks at very different positions than observed
in the Pr K-edge data.  Similarly, the Cu K-edge data for YBCO (Fig.
\ref{fig6})  also has a very different environment.  This indicates
that most of the Pr substitutes at the Y site.

We can also compare the Pr K-edge XAFS for the first Pr-O peak to the
Ce K-edge XAFS for the first Ce-O peak in CeO$_{2}$.  Both sites
should be 8-fold coordinated, and since Ce and Pr are neighbors in the
periodic table, their XAFS amplitudes and phases should be comparable.
However,  the FT of k$\chi$(k) shows a 60\% drop in amplitude in the
Pr-O peak in the 100\% Pr sample when compared to the Ce-O peak (Fig.
\ref{fig3}).  In fact, the Pr-O peak amplitude decreases with
increasing concentration (Fig.  \ref{fig5}).  The Pr-Cu peak in PBCO
shows little change with concentration, suggesting that the
praseodymium is occupying a well ordered site in the unit cell while
the oxygen is not.

Changes in the oxygen environment around the yttrium in the Y K-edge
XAFS, if any, are below the resolution of the experiment
(Fig. \ref{fig4}).
Changes in the further neighbor peaks are only evident in the Y-Ba,
Y-Y, Y-Pr regime where the double hump at 3.6  {\AA} becomes less
lopsided toward the lower side with increasing Pr concentration.
There  is no visible changes in the Y-Cu peak.

The Cu K-edge XAFS of PBCO show a 10\% reduction in the oxygen peak
amplitude compared to YBCO indicating that some disorder in the oxygen
environment exists (Fig.  \ref{fig6}).  Changes in the Cu-Ba, Cu-Y,
Cu-Pr multi-peak (between 2.6  {\AA} and 3.4  \AA) are complicated,
with the large peak at 3.15  {\AA} breaking into two humps at 3.0
{\AA}  and 3.3  \AA.  The peak due to the [Cu(1)-Cu(2)]$_{c-axis}$ and
the ab plane Cu-Cu bonds (centered at about 3.65  \AA) decreases
monotonically with Pr concentration.  This multi-peak includes the
forward scattering due to the co-linear or nearly co-linear oxygens
along most of these copper-copper paths.  Since the forward scattering
amplitude is strongly dependent on small deviations from co-linearity,
the changes in this peak probably indicate small displacements of the
O atoms.

\subsection{Background features in high-energy K-edge XAFS} A
preliminary FT of the Pr K-edge data which uses a simple spline or
polynomial for $\mu_{o}$ indicates an anomalous hump near 1 \AA.  By
fitting the data above 6 \AA$^{-1}$ to the Pr-O and Pr-Cu peaks, and
then subtracting the extrapolated fit from the original
data, clear features in
$\mu_{o}$(E) emerge
which cannot be reconciled as an XAFS signal (Fig.  \ref{fig7}).  The
position of the feature roughly 115 eV above the absorption edge is
consistent with a multielectron excitation, as estimated from the Z+1
model, i.e.,  with the N$_{IV,V}$ transition of Nd.  As pointed out by
Holland {\it et al.}, \cite{Holland78} a Ramsauer-Townsend-like
effect can also lead to significant structure in the atomic background
absorption.  Such background features arise when the wavelength of the
photoelectron is comparable to the dimension of an ``embedded'' absorbing 
atom in a solid and are essentially an XAFS of atomic origin.  The
effect appears to be larger in atoms of higher atomic number.  Calculations
by J. Rehr (private communication) show that this structure can occur
at energies comparable to those of possible multielectron excitations.
The background structure for Pr K-edge data is indeed much larger than
features we have seen in previous work.\cite{Li92}  We have removed
the background structure using the iterative procedure discussed in
Ref. \onlinecite{Li92}.
Similar features were removed from the Ce K-edge data for the standard
compound CeO$_{2}$.  Our work on these background features
will be the subject of a future paper.

\section{Detailed analysis and fit results}
\label{fits}
\subsection{Fitting procedures and constraints} All fits to the XAFS
data are carried out in r-space on the FT of k$\chi$(k).  We typically
choose an appropriate range in k for the transform, and only fit it in
some desired range in r.  In the r-space fitting procedure, we vary
the XAFS from standard compounds in amplitude, a Debye-Waller-like
broadening $\sigma$, position, and the K-edge threshold energy E$_{o}$
to obtain a good fit.  We have developed experimental standards for
many pairs of atoms.  For pairs which we have no experimental
standards, the theoretical standards calculated by FEFF5\cite{FEFF5}
were used.  Fits to the Pr K-edge data used the theoretical standards,
while Cu K-edge data were fit with experimental standards.  Since we
did not obtain standard compounds containing Pr, we evaluated the
quality of the theoretical standards for Pr by comparing the Ce-O
standard generated by FEFF5 to experimental data obtained from
CeO$_{2}$, with good agreement for the Ce-O peak
(Fig. \ref{fig3}b).
The non-linear fitting routine minimizes a fit parameter, C$^{2}$ =
$\sum$ $|$FT(k$\chi$(k)) - fit$|^{2}/<|$FT(k$\chi$(k))$|^2>$, which is
roughly proportional to the statistical $\chi^{2}$.  Each fit can
include some constraints, such as setting the number of neighbors in
one peak equal to some fraction of the number of neighbors in another
peak. Such constraints are necessary to restrict the number of
independent parameters\cite{Lin91} and can be used to test a model, or
to control correlations between fit parameters.  For instance,
correlations between amplitude and broadening factors can be
controlled by, say, constraining the sum of the number of neighbors
for two peaks to be some fixed value.  The constraints are generally
self-consistent, i.e.,  various atomic pair distances and number of
neighbors are related and must have the same results for measurements
at
different K-edges.

\subsection{Determining substitution site of Pr} A fit to FEFF5
theoretical XAFS standards that includes a mix of Pr at the Y site and
at the Ba site indicates that, at most, the Pr exists on the Ba site
8\% of the time.  This non-zero result could only be obtained if the
fit allowed the energy of the Pr K-edge (E$_{o}$) to be significantly
different (10 eV) for Pr in the Y site and Pr in the Ba site.  Such a
small concentration of Pr at the Ba site in these fits is consistent
with all Pr substituting onto the Y site.  More information regarding
this issue from the other K-edges is difficult to obtain, because the
backscattering amplitudes of both the Pr (Z=59) and Ba (Z=56) atoms
are
very similar.

A similar fit that allows some Pr to reside on the Cu(2) site is not
as simple to interpret.  For this discussion, we will refer to Pr at
the Cu(2) site as Pr$_{Cu(2)}$, and Pr at the Y site as Pr$_{Y}$.  We
can obtain a fit to the Pr K-edge data that includes 20\% of the Pr
residing on the Cu(2) site, with a lengthening of the Pr$_{Cu(2)}$-O
bonds by 0.25  {\AA} and of the Pr$_{Cu(2)}$-Pr$_{Y}$ bonds by 0.13
\AA.  Such a distortion in the oxygen environment is consistent with
both the Pr K-edge and the Cu K-edge data.  However, a distortion in
the Pr$_{Cu(2)}$-Pr$_{Y}$ bond should effect how the Pr$_{Y}$ sits in
the unit cell, and should thus be reflected in the Pr$_{Y}$-Cu bond.
This  effect should be large even for small ($\sim$ 20\%) Pr$_{Cu(2)}$
concentrations, because the shift in the Pr$_{Cu(2)}$-Pr$_{Y}$ bond
length is large enough to cause nearly a complete cancellation of the
distorted Pr$_{Y}$-Cu peaks.  The data (Fig.  \ref{fig5}) and the fits
(see Sec.  \ref{levelPr}) show very little change in the Pr-Cu peak.
Therefore,  these data are not consistent with Pr at the Cu(2) site
within the amplitude limits of XAFS spectra ($\sim$ 10\%), for all Pr
concentrations
measured.

\subsection{Y K-edge fits}
\label{levelY}
Fits to the Y K-edge data were performed, without any constraints, on
the 0, 30, and 50\% Pr samples.  Results for the oxygen and copper
peaks are in Table \ref{table2}.  Further peaks, such as the Y-Ba,
Y-Pr, and the Y-Y peak were fit but are not reported, because they are
all at similar distances, and the Y-Ba, Y-Pr peaks are almost exactly
out of phase with the Y-Y peak.  This situation allows changes in
amplitude to be mimicked by changes in position, and thus does not
give reliable results.

The number of oxygen neighbors shown in Table \ref{table2} changes
very little with concentration (about 5\% up to 50\% Pr), while the
copper peak shows a slight increase.  Such small changes are not
within the resolution of this experiment, and the data are therefore
consistent with no change in the number of O atoms surrounding Y.  The
Y-O distance remains constant.  The Y-Cu distance does increase
monotonically with Pr concentration, as one would expect from the
diffraction results
(Table \ref{table1}).

\subsection{Pr K-edge fits to a variety of models}
\label{levelPr}
Fits to the Pr K-edge XAFS spectra were performed with and without a
variety of constraining models and initial conditions.  Each model has
an associated C$^{2}$ value for its fit, which should give the reader
a feel for the range of models that may give plausible results.  Table
\ref{table3} gives fits to the 100\% Pr sample with a variety of
constraints imposed on the fits that simulate different models.  None
of the models constrain the Pr-Cu peak, and none of the fits to these
models show any significant differences in this
peak.

Model 1 assumes a single Pr-O bond, with no constraints, and gives a
good quality of fit.  This fit shows an oxygen deficiency around Pr,
with only 4.8  of the expected 8 oxygens.  Since this result is
inconsistent with the fit results of the Y K-edge (Sec.
\ref{levelY}),  the Cu K-edge (Sec.
\ref{fits,levelCu}) and the
volumetric measurements of the oxygen concentration in these samples
(Table \ref{table1}), we constrained the next fit to give 8 oxygens.
Model 2, which
has one Pr-O distance and does not allow for an oxygen deficiency,
yields a poor fit with a substantially larger $\sigma$ and an increase
in the goodness-of-fit parameter (C$^{2}$) by more than an order of
magnitude. It is therefore highly unlikely that simple harmonic
disorder of the Pr-O bond can explain the reduction in the Pr-O peak
in the FT of k$\chi$(k).

   Next we considered
several models that allow the first peak to be composed of two
different
Pr-O harmonic distributions.  
Since for K-edge data any shift due to differences in valence should
be small, all of these models constrain any E$_{o}$ shifts in the Pr-O
bonds to be the same.  Model 3 allows for a double Pr-O bond, with no
further constraints.  This fit has the best C$^{2}$ of any of the
models. It indicates a total number of oxygens which is still low
(5.7). However, a fit to model 4, which constrains the total number of
nearest-neighbor oxygens to sum to 8, gives an C$^{2}$ which is only
30\% higher.

Two other models were tested to see if the fits in models 3 and 4 can
be distinguished from similar models.  All these fits have a good
quality of fit parameter, but shift certain parameters outside a
reasonable range.  A fit which constrains the length of the two bonds
to be equal and the total number of oxygens to be 8 (model 5), but
allows for different $\sigma$'s gives a very large $\sigma$ of 0.287
{\AA}  for the short Pr-O bond, with 3.5  O neighbors (roughly 36\% of
the O).  However, a peak this broad has a tiny XAFS amplitude, and
therefore this fit is essentially the same as the single Pr-O peak fit
in model 1.  Consequently  we do not consider this possibility further.
Constraining  the $\sigma$'s to be equal (rather than the bond
lengths), with 8 total oxygen neighbors (model 6) gives a fit with a
short Pr-O bond length of 2.15  {\AA} and an increase in C$^{2}$ by
less than a factor of two over model 4.  In this fit, only 15\% of the
Pr-O bonds are short.  Such a short bond length is not unreasonable,
given that the short Pr-O bond in BaPrO$_{3}$ is 2.18
{\AA},\cite{Wyckoff}  although Pr is only 6-fold coordinated with
respect to oxygen in that compound.  Models 3, 4, and 6 all indicate
that the radial distribution of Pr-O bond lengths is not a symmetric
distribution; the major weight is near 2.45  {\AA} but there is
clearly some weight in a broad peak centered near
2.27$^{+0.03}_{-0.12}$ {\AA}.

Fits that allow for three Pr-O distributions (as suggested by Ref.
\onlinecite{Fehrenbacher93})  are not conclusive, because the number
of parameters required for such a fit exceeds the maximum number of
parameters allowed by our fitting range.\cite{Lin91}  However, the two 
peak fits described above giving large $\sigma$'s for the shorter Pr-O
distribution may mimic a situation with three distances.

The peaks that allow for two Pr-O distributions give the most
consistent results.  Therefore, fits that allow the Pr-O peak to be
composed of two different harmonic distributions (models 3 through 6)
suggest 15-40\% of the nearest-neighbor Pr-O bonds are either
disordered and/or shifted ($\simeq$ 0.18  \AA) compared to a more
populated, ordered site.

Table \ref{table4} summarizes the results of a two-peak fit with the
constraint that there be 8 oxygens in the first peak for all measured
Pr concentrations.  Debye-Waller factors for the disordered site
increase linearly with Pr concentration, and are as much as 150\%
higher than for the ``normal'' site in PBCO.  The fits show no
appreciable change in the Cu environment around praseodymium with
concentration, apart from the expected lengthening of the Pr-Cu bond
due to the slight expansion of the lattice (Table \ref{table1}).  The
number of copper neighbors around the praseodymium remains the same
for the 100\% Pr sample compared to the other Pr concentration
samples. Therefore, we can conclude that the effect of the small
fraction of other phases in the PBCO sample \cite{Zhu92} on the XAFS
is below the resolution of the experiment.

\subsection{Cu K-edge fits}
\label{fits,levelCu}
Cu K-edge fits to the 0\% and 100\% Pr samples were also carried out.
Fits  to partial concentrations of Pr are difficult to obtain
because of the
known distortions caused by the two sub-lattices.  Several constraints
were placed on the fits to maintain certain symmetries in the system.
For  instance, the Cu(1)-O(4) distance plus the Cu(2)-O(4) distance is
constrained to equal the Cu(1)-Cu(2) distance obtained from
diffraction.\cite{Lowe-Ma92}
The Ba
position is constrained to be in the center of the ab-plane of the
unit cell, while the Y or Pr position is constrained to be in the
middle of all planes in the unit cell, consistent with the Cu(2)-Y and
the Cu(2)-Cu(2) (c-axis) bond lengths.  Other constraints on the
number of neighbors have been imposed to maintain the number of atoms
at certain sites while still allowing for overall shifts in amplitude.
The  proximity of the Cu(1)-O(1), Cu(2)-O(2), and Cu(2)-O(3) peaks
makes an accurate determination of their relative amplitudes
difficult. Consequently, measurements of changes in their relative
amplitudes is beyond the resolution of our fits.  We have treated the
Cu(1)-O(1), Cu(2)-O(2), and the Cu(2)-O(3) as one peak (the
Cu-O$_{planar}$ peak), and constrained the relative amplitudes of this
peak to the Cu(1)-O(4) and the Cu(2)-O(4) peaks.

The fits to the Cu K-edge data are summarized in Table \ref{table5}.
Fit  parameters for YBCO are typical of fits we have obtained in
previous work.  There are several differences between the YBCO and the
PBCO results.  Bond length changes are, of course, expected, and will
be dealt with in some detail in the next section.  Number of neighbor
values are 15\% different for the combined Ba and Pr peak compared to
YBCO, which is usually considered to be outside the expected error
(10\%) for XAFS amplitudes.  However, the amplitude of this combined
peak is very sensitive to the individual positions of the Ba and the
Pr, as well as to the position of the Cu(2)-Cu(2)$_{c-axis}$ peak, and
are therefore not very reliable.  The broadening factors are slightly
larger for the planar oxygens and the Cu(1)-O(4) bond in PBCO, as are
the $\sigma$'s for the Ba and Pr peaks.  The Cu(2)-O(4) bond in PBCO
is much narrower than in the YBCO.  Again, the $\sigma$'s for the Ba
and Pr peak suffer from the same problems as their amplitudes, but the
change in $\sigma$ for the oxygen peaks is probably real.

\section{Discussion}
\label{discussion}

\subsection{Consistency between various K-edge data}

Since each absorption edge gives us information about the oxygen
environment within YBCO:Pr, we can constrain our results to give a
consistent picture.  Y K-edge data show little if any change in the
structure of YBCO:Pr up to 50\% Pr, except (possibly) in the expected
lengthening of the Y-Cu bond.  Even this expansion of the Y-Cu
distance is at the edge of the accuracy of this experiment.  These data
therefore show that up to 50\% Pr, the oxygen and copper environment
around yttrium in YBCO: Pr is essentially the same as in pure YBCO.
More  specifically, there is little evidence of a loss of oxygens or
of disorder in the oxygen or copper environment.

The Pr K-edge data and fit results indicate that either there are
missing oxygens in the Cu-O planes or that there is a large amount of
disorder and/or distortion in these planes.  The former is
inconsistent both with the measured oxygen content and with the number
of neighbors for the Cu-O and the Y-O peaks in the XAFS data, and is
therefore ruled out.  The data is best fit by two Pr-O distributions,
separated in pure PBCO by about 0.18  {\AA}, with one distribution
less populated and more disordered than the other.

The Cu K-edge data is more complicated because of the two copper
sites. Nevertheless, the oxygen environment as seen from the coppers
should be consistent with (and therefore can help constrain) the Pr
K-edge results.  The fit results to the oxygen peaks show essentially
no change in the number of nearest-neighbors, but indicate a
broadening in the Cu-O$_{planar}$ peak consistent with the 10\%
decrease in the peak height.  If the Pr-O distortion is completely
radial, then a shortening of 0.18  {\AA} would correspond to only a
0.04 {\AA} lengthening of $\sim$30\% of the Cu(2)-O(2,3) bonds, which
is not
resolvable in the
fits.  However, such a distortion will contribute to a broadening of
the Cu- O$_{planar}$ peak, and therefore the measured broadening is
consistent with the proposed radial distortion in the Pr-O bond.  The
small change in the number of Y-O neighbors (which can also be modeled
as a slight increase in $\sigma$) is also consistent with a radial
distortion in the Pr-O bonds.

\subsection{On the question of valence}

The fit results to the first oxygen peak from all three edges are
consistent with a distortion of the Pr-O bond in the radial direction
of approximately 0.18  \AA.  This distortion may be due simply to a
hybridization of the O 2{\it p} and Pr 4{\it f} electrons.  However,
the magnitude of the shift is very consistent with some Pr$^{4+}$; the
best fit gives a Pr-O$_{short}$ bond length (2.27  \AA) which is close
to the nearest-neighbor bond length in 8-fold coordinated PrO$_{2}$.
By  constraining the broadening factors, we can obtain a much shorter
bond length (2.15  \AA) which is more consistent with 6-fold
coordinated BaPrO$_{3}$.  The comparison to the 8-fold coordinated
PrO$_{2}$ makes the most sense, since Pr in PBCO is 8-fold
coordinated, although it is possible that not all the oxygens
participate in
Pr-O bonds.  

If we assign all of the short Pr-O bond lengths to the formal
Pr$^{4+}$ state, our fit results indicate that the Pr is in that state
15-40\% of the time, corresponding to a formal valence of
+3.33$^{+0.07}_{-0.18}$. Such an assignment is reasonable for the
shortest measured Pr-O bond length, but if the bond length is closer
to $\sim$2.3  {\AA}, then some could still include Pr$^{3+}$, as in
bixbyite Pr$_{2}$O$_{3}$.  Therefore, the upper limit on the valence
range may be too high.  Likewise, the lower limit was obtained by
forcing the broadening factor of the short bond to be much lower than
in the best fit, which may not be reasonable.

The possible disorder in the shorter Pr-O bond could cause
localization of charge carriers, and could be a byproduct of a 4{\it
f}-2{\it p} hybridization.  On the other hand, since the disorder is
greatest for the PBCO sample, the disorder could be due to the less
common Pr$^{4+}$-Pr$^{4+}$ combination, which will probably have
different Pr-O bond lengths than either the Pr$^{3+}$-Pr$^{3+}$ or the
Pr$^{3+}$-Pr$^{4+}$ combinations.

Our measured dominant Pr-O distance of 2.45$\pm$0.01  {\AA} is in
excellent agreement with diffraction
results\cite{Morales90,Neumeier90,Lowe-Ma92,Guillaume93} that give a
mean Pr-O distance of 2.4539$\pm$0.0038  \AA.  We see no evidence of a
longer distance ($\simeq$ 2.5  \AA) as suggested by the Pr-O bond
length trend with ionic radius.\cite{Guillaume93}.  Fits that include
such a peak invariably shift the peak back to the values we report.
Therefore,  we must conclude that the discrepancy of this bond length
with ionic radius is not directly due to a mixed valency.  Indeed,
given the existence of a short Pr-O bond, the long Pr-O bond seems to
be entirely Pr$^{3+}$(4{\it f}$^{2}$)-O.

The range for the valence is in approximate agreement with the
electronic studies mentioned in Sec.  \ref{introduction} that give the
valence of Pr to be close to 3+.  However our results clearly suggest
a higher formal valence than 3.0,  as was found in Pr L-edge studies
of Horn {\it et al.}  (3.1+  for PBCO)\cite{Horn87} and Lytle {\it et
al.} (3.45+  for 20\% Pr and 3.25+  for 60\% Pr).\cite{Lytle90}  In
addition, our structural results show that the number of short Pr-O
bonds, that is, formal Pr$^{4+}$, increases with increasing Pr
concentration, and that any Pr$^{4+}$ that may be present resides on
the Y site.  Both of these results are in contrast to the Lytle {\it
et al.}  results.

Our result is in striking agreement with the 4+ / hybridization model
put forth by Fehrenbacher and Rice,\cite{Fehrenbacher93} which
predicts that the Pr$^{3+}$ state will exist 60-70\% of the time.  A
ligand hole localized on the Pr-O(2)/O(3) bond
would remove a
hole from the conduction band and yield a formal Pr$^{4+}$ site.  The
resulting Pr$^{4+}$-O bond length would likely be significantly
shorter than the Pr$^{3+}$-O bond, based on the observed bond lengths
in other materials.  Although we cannot give a precise estimate of the
number of short bonds present, our results are consistent
with their prediction.

\subsection{On the question of the relative position of O(4)}

The Cu K-edge data for PBCO are consistent with the diffraction
results of Lowe-Ma and Vanderah \cite{Lowe-Ma92} which show the axial
oxygen (O(4)) to be pushed toward the planar Cu site when compared to
the YBCO structure (Table \ref{table4}).  The comparison to YBCO is
properly made by taking the differences in the lattice parameters a,
b, and c into account.  Therefore, we also show the bond-lengths for
``pseudo-YBCO'' which is calculated using the lattice parameters for
PBCO from Lowe-Ma and Vanderah\cite{Lowe-Ma92} and the relative atomic
positions of YBCO from Beno {\it et al.}\cite{Beno87}  The difference
in Cu(2)-O(4) distance between the diffraction and the XAFS results
can be explained by comparing the Cu(2)-O(4) distance as a function of
oxygen concentration for R123-type materials.  In all these materials
that have been measured, the Cu(2)-O(4) distance decreases with
increasing oxygen concentration.  Our result is consistent with this
trend (see Ref.  \onlinecite{Lowe-Ma92}).  An extrapolation of the
measured Cu(2)-O(4)
distance as a function of oxygen concentration in
PBCO,\cite{Lowe-Ma92} to an O content of 6.98,  yields a bond length
of 2.23  {\AA}, in excellent agreement with our result of 2.22  {\AA}.
 
\subsection{On the question of clustering in RBCO:Pr (R=rare earth and
Y)}

An interesting property of R$_{1-x}$Pr$_{x}$Ba$_{2}$Cu$_{3}$O$_{7}$ is
that the larger the rare-earth radius, the lower the Pr concentration
required to completely suppress T$_{c}$.  It has been suggested
\cite{Zhu92} that this is evidence for clustering of Pr in YBCO:Pr
(where the ionic-size difference is comparatively large), thereby
maximizing the superconducting Y-Y pairs of cells within a crystal.
We  had expected that both the Pr and Y K-edge data would be very
sensitive to this clustering, by giving the number of neighboring Y
and Pr atoms.  The data in Fig.  \ref{fig4} and \ref{fig5} show a
significant change in the structure near 3.5  {\AA} where the Y or Pr
neighbor peak occurs, compared to the pure materials YBCO and PBCO.
This  means that in the 30 and 50\% samples there are both Y and Pr
neighbors at each Y/Pr site.  If large scale clustering were present,
these significant changes in the XAFS would not occur.  Unfortunately,
fits that try to precisely determine the ratio of Pr to Y neighbors
are plagued by the interference effects from the Y-Ba and Y-Y peaks,
as described in Sec.  \ref{levelY}, which reduces our sensitivity.
Consequently,  we cannot differentiate between a random distribution
and small clusters.

\section{Conclusion}
\label{conclusion}

In summary, we have collected XAFS data on samples with several
different concentrations of Pr in
Y$_{1-x}$Pr$_{x}$Ba$_{2}$Cu$_{3}$O$_{7}$ and on CeO$_{2}$ at the Y,
Pr, Cu, and Ce K-edges.  The main result of this investigation is that
the XAFS data and analysis show that disorder in the oxygen
environment around the praseodymium clearly exists.  This disorder is
most obvious when comparing the first oxygen peak in CeO$_{2}$ to the
first oxygen peak in PrBa$_{2}$Cu$_{3}$O$_{7}$, which shows a 60\%
reduction in the Pr-O peak.  The amplitude of this peak clearly
shrinks with increasing Pr concentration.

This PBCO data is best fit by two harmonic radial distributions for
the Pr-O bond, which are centered at 2.27$^{+0.03}_{-0.12}$  {\AA} and
2.45$\pm$0.01 {\AA} in PBCO.  The shorter bond may
be broader than
the longer one by as much as a factor of 3.  The magnitude of this
distortion agrees well with the change in bond length in other Pr
oxides between Pr$^{4+}$-O (2.18  {\AA} - 2.32  \AA) and Pr$^{3+}$-O
(2.33 {\AA} - 2.66  \AA) and is therefore the first structural
evidence of a split in the Pr-O bond lengths, and strong evidence for
a mixed valent state of Pr in
Y$_{1-x}$Pr$_{x}$Ba$_{2}$Cu$_{3}$O$_{7}$. 

We have also verified some results of previous measurements.  By
comparing the Pr K-edge data to FEFF5 simulations of
PrBa$_{2}$Cu$_{3}$O$_{7}$ with Pr at the Y site and at the Ba or Cu(2)
site, we see very little evidence for any of the Pr existing on the Ba
or the Cu(2) site (upper limits of 8 and 10\%, respectively).  In
addition, Cu K-edge data indicate the O(4) site has moved closer to
the Cu(2) site (in agreement with Ref.  \onlinecite{Lowe-Ma92}) and
therefore may impede charge transfer between the planes and the
chains.

\acknowledgments 
We especially wish to thank G. G. Li, J. Rehr and T.H.  Geballe for 
several
useful discussions.  The experiments were performed at the Stanford
Synchrotron Radiation Laboratory, which is operated by the U.S.
Department  of Energy, Division of Chemical Sciences, and by the NIH,
Biomedical Resource Technology Program, Division of Research
Resources. The work is supported in part by NSF, grant number
DMR-92-05204.


\begin{figure}
\caption{ T$_{c}$ vs.  Pr concentration.  The inset shows the
structure of YBa$_{2}$Cu$_{3}$O$_{7}$ with the notation used in this
paper.
\label{fig1}}
\end{figure}

\begin{figure}
\caption{k$\chi$(k) for the 50\% Pr sample from the Y, Pr and Cu
K-edges (at 17080 eV, 41991 eV, and 8979 eV, respectively).
\label{fig2}}
\end{figure}

\begin{figure}
\caption{
Fourier Transforms of k$\protect\chi$(k) for (a)
PrBa$_{2}$Cu$_{3}$O$_{7}$ and (b) CeO$_{2}$, together with fits to the
first two major peaks (dotted).  The outer envelope is the magnitude 
of the FT, while the modulating curve is the real part.  
All the fits discussed in the text
are of similar quality to the PBCO fit shown here.  The first peak in
both cases is an eight-fold coordinated oxygen peak.  Single pair peak
positions are approximately given by vertical lines that meet the
x-axis. These peak positions are shifted from the actual positions by
phase shifts at the absorbing and backscattering atoms.  Both
transforms are from 3.5-17.0  \AA$^{-1}$, with a 0.3  {\AA$^{-1}$}
gaussian window.
\label{fig3}}
\end{figure}

\begin{figure}
\caption{Fourier Transform of k$\chi$(k) for Y K-edge data for 0\%
(solid), 30\% (dotted), and 50\% (dash) Pr concentrations.  The only
dramatic changes occur in the Y-Ba, Y-Y, Y-Pr region.  These
transforms are taken from 3.5-15.5  \AA$^{-1}$, with a 0.1
{\AA$^{-1}$}  gaussian window.
\label{fig4}}
\end{figure}

\begin{figure}
\caption{Fourier Transforms of k$\chi$(k) for Pr K-edge data for 30\%
(solid), 50\% (dotted), and 100\% (dash) Pr concentrations.  Notice
the decrease of the first peak with Pr concentration.  This decrease
is most likely due to a combination of the relative amplitudes of two
Pr-O distances and additional disorder in the shorter bond.  The
second peak is the Pr-Cu peak.  It shows relatively little change with
Pr concentration, indicating that the Pr is well ordered with respect
to the Cu, and thus to the unit cell.  The peaks in the 3-4 {\AA}
range are due to a mix of Pr-Y, Pr-Pr and Pr-Ba.  This region is
difficult to fit accurately because the Y and Pr backscattering
amplitudes are similar, but $\pi$ out of phase.  The FT ranges are
from 3.5-17.0  \AA$^{-1}$, with a 0.3  {\AA$^{-1}$} gaussian window.
\label{fig5}}
\end{figure}

\begin{figure}
\caption{Fourier Transforms of k$\chi$(k) for Cu K-edge data for 0\%
(solid), 30\% (dotted), 50\% (short dashed), and 100\% (long dashed)
Pr concentrations.  The first Cu-O peak decreases with increasing Pr
concentration by about 10\% in PBCO.  The Cu-Y, Cu-Pr, Cu-Ba region
also changes dramatically, due to the changing phase shifts between
these peaks; see text for further discussion.
These transforms
are taken from 3.0-15.8  \AA$^{-1}$, with a 0.1  {\AA$^{-1}$} gaussian
window.
\label{fig6}}
\end{figure}

\begin{figure}
\caption{``Free-atom absorption'' $\mu_o$ (solid) and total absorption
$\mu$ (dotted) coefficients times sample thickness, t, as a function
of energy above the
K-edge.  
$\mu_o$t was determined by an iterative procedure.\protect\cite{Li92}
\label{fig7}}
\end{figure}

\onecolumn
\narrowtext

\begin{table}
\setdec 00.000
\caption{Sample information: a, b, and c lattice parameters for the
Y$_{1-x}$Pr$_{x}$Ba$_{2}$Cu$_{3}$O$_{y}$ samples.}
\begin{tabular}{ccccc}
x&a&b&c&y\\
\tableline
0.0&3.821&3.882&11.621&6.96\\
0.3&3.826&3.894&11.667&6.94\\
0.5&3.846&3.907&11.691&6.96\\
1.0&3.902&3.916&11.715&6.98\\
\end{tabular}
\label{table1}
\end{table}

\mediumtext
\begin{table}
\setdec 00.000
\caption{Y K-edge fit results.  These fits include further peaks than
the Y-Cu peak, but these peaks are plagued by an interference effect
when Pr is present (see Sec.  \protect\ref{levelPr}).  Y-Cu amplitudes
are consistently larger than those calculated by FEFF5 by nearly 25\%.
Since  we do not have data for a similar standard pair, we have
normalized the Y-Cu number of neighbors to the YBCO result.  The
number of Y-O nbrs is not normalized.  The fits are from 1.0-3.1
{\AA}  in R, and from 3.5-17  {\AA$^{-1}$} in the wave vector k.  Each
bond is expected to be comprised of 8 neighbors.  The errors are
$\pm$0.01 {\AA} in R, and $\pm$10\% in both $\sigma$ and number of
neighbors for all XAFS fits reported in this work, unless otherwise
noted.}
\begin{tabular}{cccccccc}
&\multicolumn{3}{c}{Y-O}&&\multicolumn{3}{c}{Y-Cu}\\
x&R(\AA)&nbrs&$\sigma$(\AA)&&R(\AA)&nbrs
&$\sigma$(\AA)\\
\tableline
0.0&2.41&7.3&0.058&&3.21&8.0&0.058\\
0.3&2.41&7.1&0.056&&3.22&8.6&0.060\\
0.5&2.41&6.9&0.057&&3.23&8.6&0.060\\
\end{tabular}
\label{table2}
\end{table}

\widetext
\begin{table}
\setdec 00.000
\caption{Fit results on PrBa$_{2}$Cu$_{3}$O$_{7}$ to a variety of
models. Fitting procedures are detailed in the text.  C$^{2}$ is the
fitting parameter, and is approximately proportional to the
statistical $\chi^{2}$, as described in the text.  The r-space fits
are from 1.0  - 3.0  {\AA} in R, and the FT from 3.5-17.0  \AA$^{-1}$
in wave vector k.  The number of Pr-O neighbors is normalized to the
CeO$_{2}$ result.  No normalization has been applied to the Pr-Cu
peak. Model 1 only allows for one Pr-O distance, with no constraints
on the fit parameters.  Model 2 also only allows for a single Pr-O
bond, but constrains the amplitude to be 8, as is expected from the
known crystal structure.  Model 3 allows for two Pr-O bonds (Pr-O$
_{short}$ and Pr-O$_{long}$, as do all the subsequent models) with no
further constraints.  Model 4 constrains the total number of Pr-O
bonds to 8.  Model 5 constrains the number of Pr-O bonds to 8 and
holds their bond lengths
equal.
Model 6 constrains the total number of Pr-O bonds to 8 and holds their
broadening factors ($\sigma$) equal.}
\begin{tabular}{cccccccccccccc}
&\multicolumn{3}{c}{Pr-O$_{short}$}&&\multicolumn{3}{c}{Pr-O$_{long}$}
&&\multicolumn{3}{c}{Pr-Cu}\\
model&R(\AA)&nbrs&$\sigma$(\AA)&&R(\AA)&nbrs&$\sigma$(\AA)&&
R(\AA)&nbrs&$\sigma$(\AA)&&C$^{2}$\\
\tableline
1&&&&&2.46&4.8&0.066&&3.27&7.8&0.049&&0.148\\
2&&&&&2.47&8.0&0.116&&3.27&7.6&0.047&&1.513\\
3&2.30&1.0&0.135&&2.46&4.7&0.063&&3.27&7.9&0.048&&0.132\\
4&2.27&2.9&0.168&&2.45&5.1&0.067&&3.27&7.7&0.048&&0.174\\
5&2.46&3.5&0.287&&2.46&4.5&0.064&&3.27&7.8&0.048&&0.160\\
6&2.15&1.2&0.084&&2.44&6.8&0.084&&3.27&7.7&0.048&&0.300\\
\end{tabular}
\label{table3}
\end{table}

\begin{table}
\setdec 00.000
\caption{Pr K-edge fit results to
Y$_{1-x}$Pr$_{x}$Ba$_{2}$Cu$_{3}$O$_{7}$ as in model 4 in Table
\protect\ref{table3}.}
\begin{tabular}{cccccccccccc}
&\multicolumn{3}{c}{Pr-O$_{short}$}&&\multicolumn{3}{c}{Pr-O$_{long}$}
&&\multicolumn{3}{c}{Pr-Cu}\\
x&R(\AA)&nbrs&$\sigma$(\AA)&&R(\AA)&nbrs&$\sigma$(\AA)&&
R(\AA)&nbrs&$\sigma$(\AA)\\
\tableline
0.3&2.25&1.7&0.128&&2.43&6.3&0.056&&3.25&7.5&0.046\\
0.5&2.29&2.7&0.115&&2.43&5.3&0.053&&3.25&7.4&0.047\\
1.0&2.27&2.9&0.168&&2.45&5.1&0.067&&3.27&7.7&0.048\\
\end{tabular}
\label{table4}
\end{table}

\begin{table}
\setdec 00.000
\caption{Cu K-edge fit results to YBa$_{2}$Cu$_{3}$O$_{7}$ (YBCO) and
PrBa$_{2}$Cu$_{3}$O$_{7}$ (PBCO).  The ``pseudo-YBCO'' is calculated
from the lattice parameters of PBCO\protect\cite{Lowe-Ma92} and the
relative atomic positions in YBCO.\protect\cite{Beno87}  The last
column shows the differences between the XAFS results and the
diffraction results for PBCO.  The samples used in the diffraction
measurements for PBCO had a mean oxygen content of $\simeq$ 6.77.  The
PBCO XAFS results agree with the
diffraction results,
within calculated errors, except for the Cu(1)-O(4) and the Cu(2)-O(4)
bond lengths, as discussed in Sec.  \protect\ref{discussion}.  The
calculated errors are the same as discussed in Table
\protect\ref{table2} except for the Cu-Ba,Pr neighbors in PBCO ($\pm$
1.5), and the positions of the Cu-Ba, Pr and Y peaks for both PBCO
($\pm$ 0.03  \AA) and YBCO ($\pm$ 0.02  \AA).
}
\begin{tabular}{lccdcccdcccc}
&expected&\multicolumn{3}{c}{PBCO XAFS}&&
\multicolumn{3}{c}{YBCO XAFS}&
pseudo-YBCO&PBCO diffrac.
\protect\cite{Lowe-Ma92}&XAFS-diffrac.\\
bond&nbrs&R(\AA)&\multicolumn{1}{c}{nbrs}&
$\sigma$(\AA)&&R(\AA) &\multicolumn{1}{c}{nbrs}&$\sigma$(\AA)
&R(\AA) &R(\AA)&R(\AA)\\
\tableline
Cu(1)-O(4)&2&1.88&2.4&0.082&&1.85&2.3&0.069&1.856&1.849&+0.031\\
Cu-O$_{planar}$&10&1.98&11.8&0.082&&1.94&11.7&0.069&1.965&1.965&+0.015\\
Cu(2)-O(4)&2&2.22&2.4&0.056&&2.27&2.3&0.110&2.311&2.254&$-$0.032\\
Cu(2)-Ba  &8&3.36&9.5&0.056&&3.37&7.6&0.035&3.408&3.388&$-$0.018\\
Cu(1)-Ba  &8&3.47&9.5&0.056&&3.49&7.6&0.035&3.499&3.482&$-$0.012\\
Cu(2)-Y,Pr&8&3.23&9.5&0.051&&3.21&7.6&0.051&3.232&3.265&$-$0.025\\
\end{tabular}
\label{table5}
\end{table}

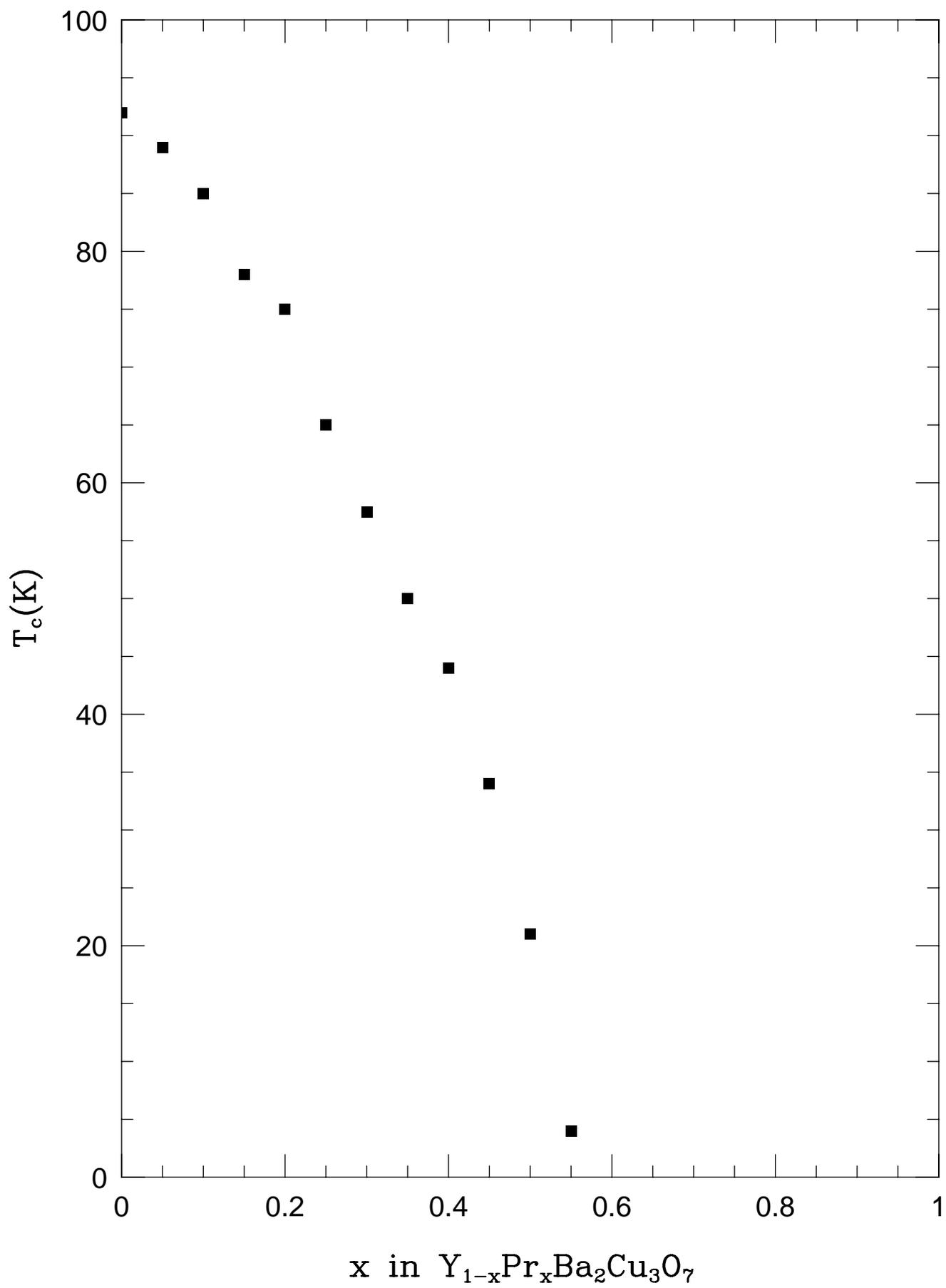
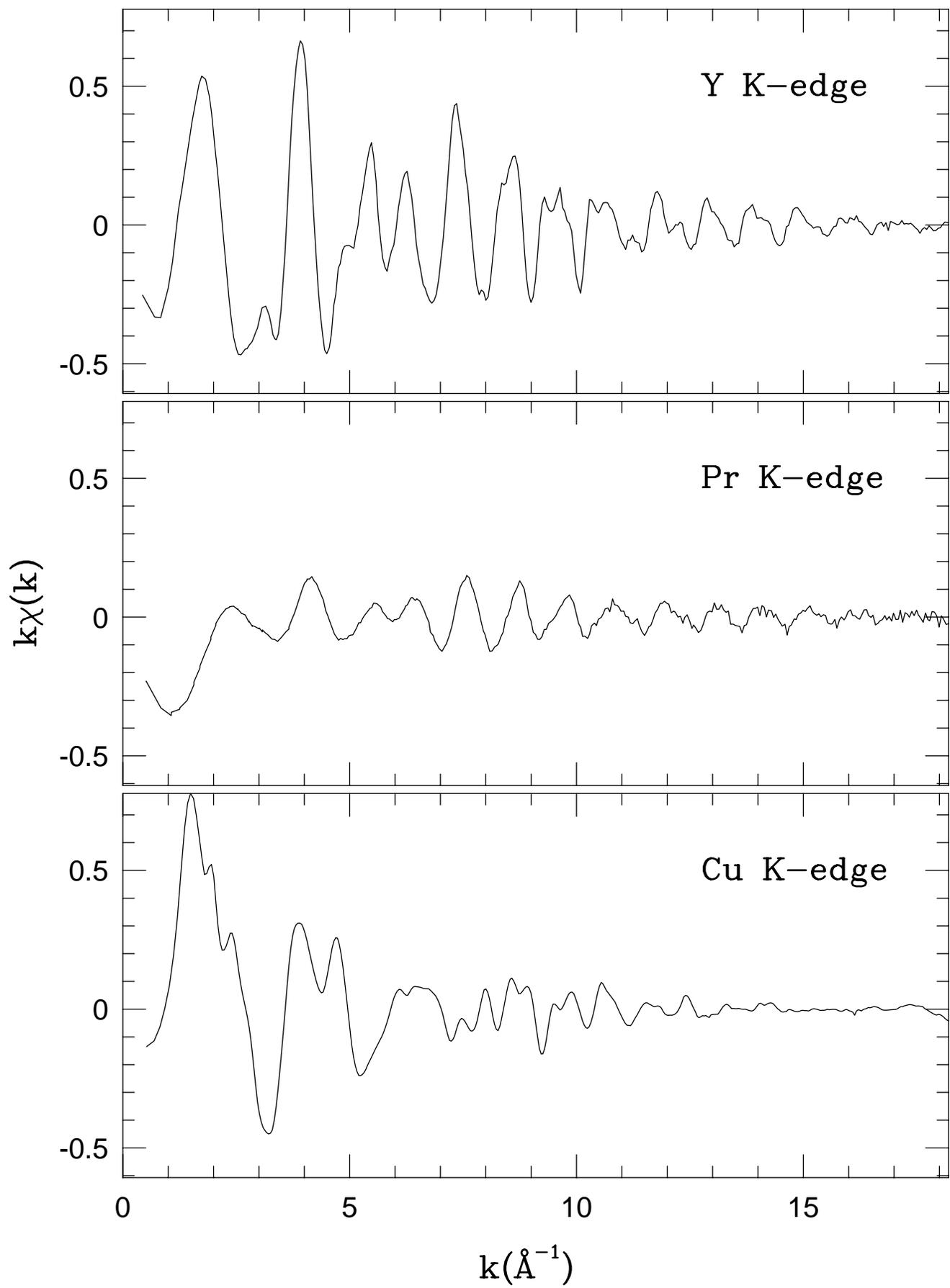
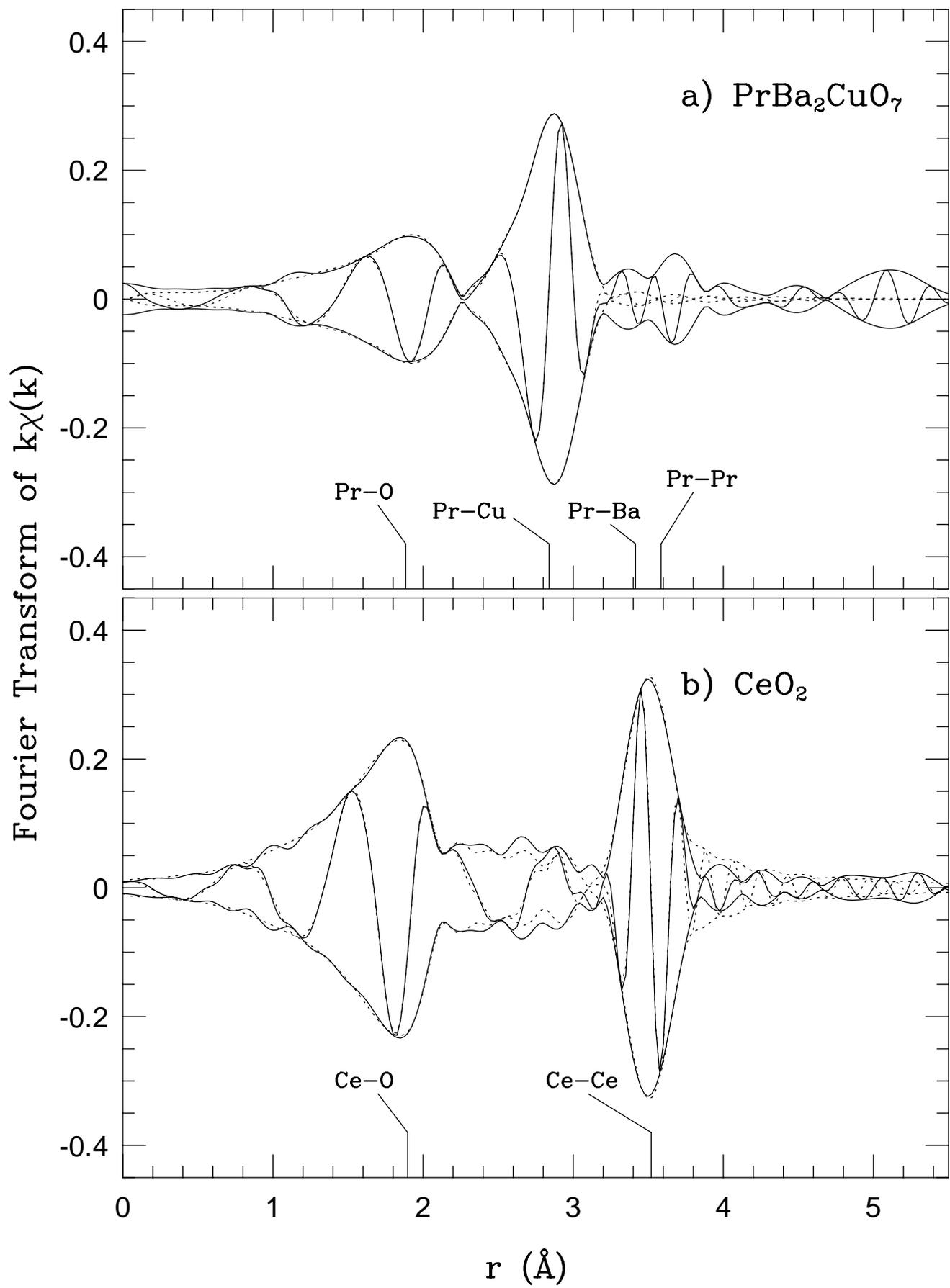
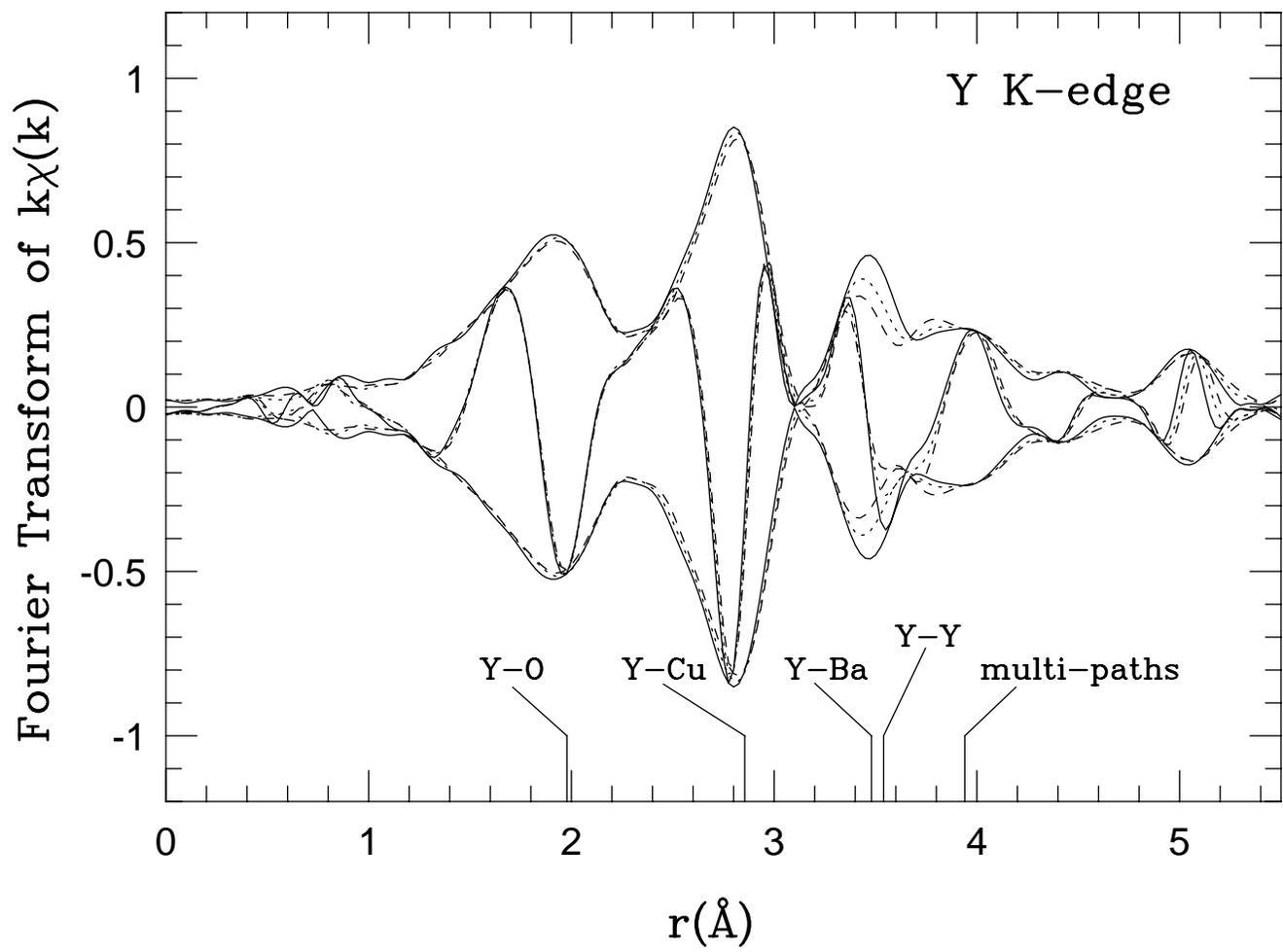
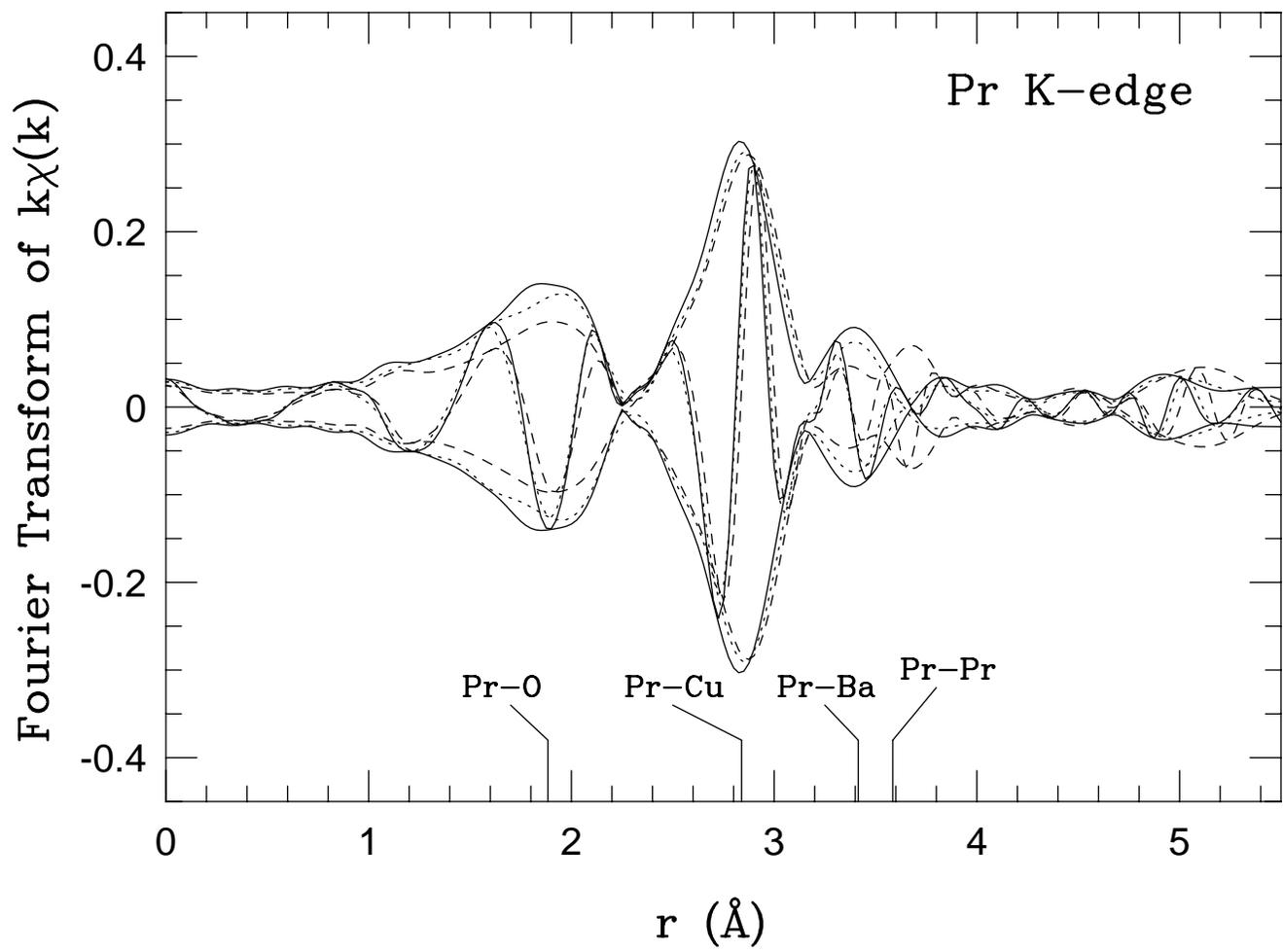
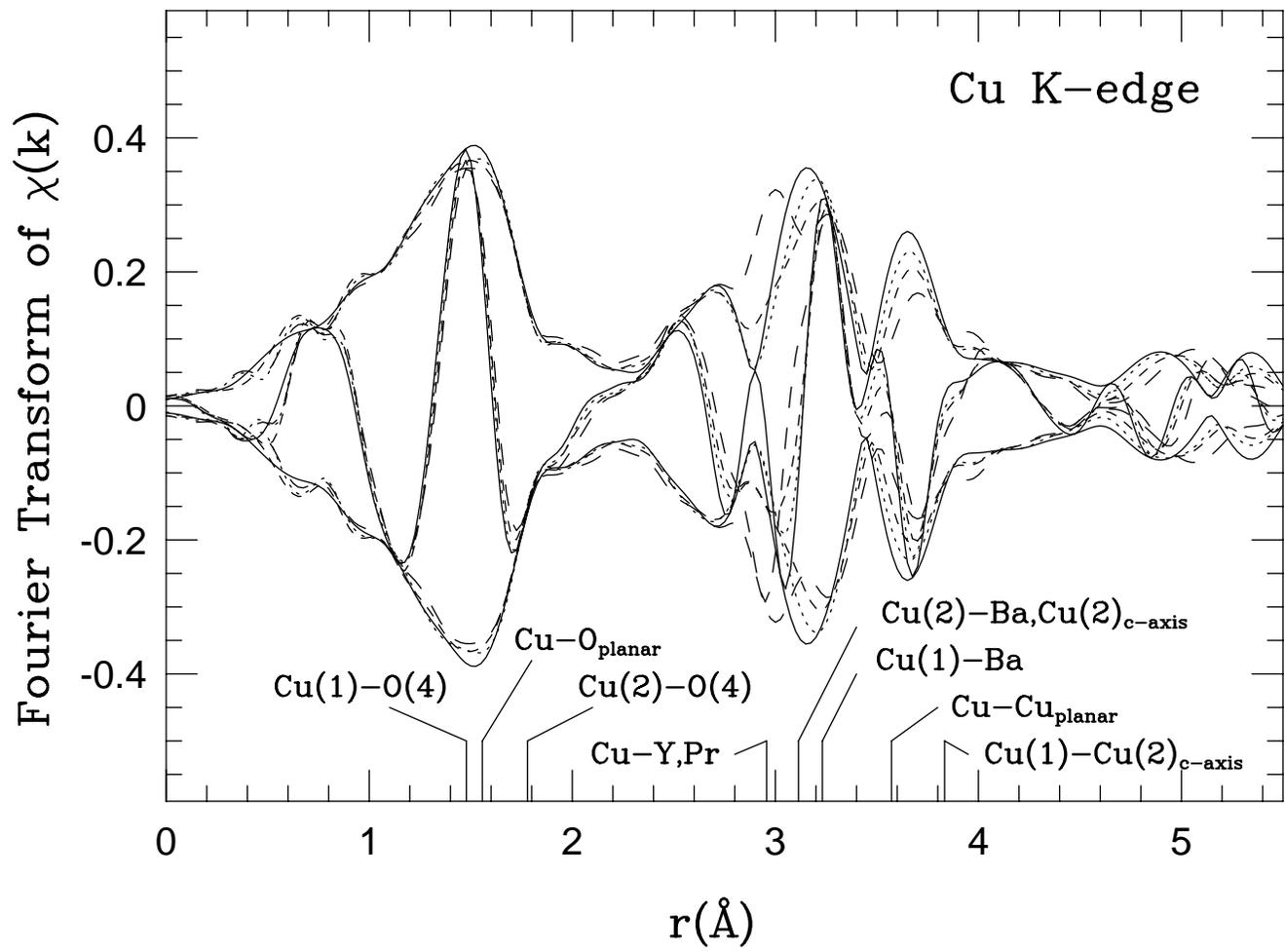
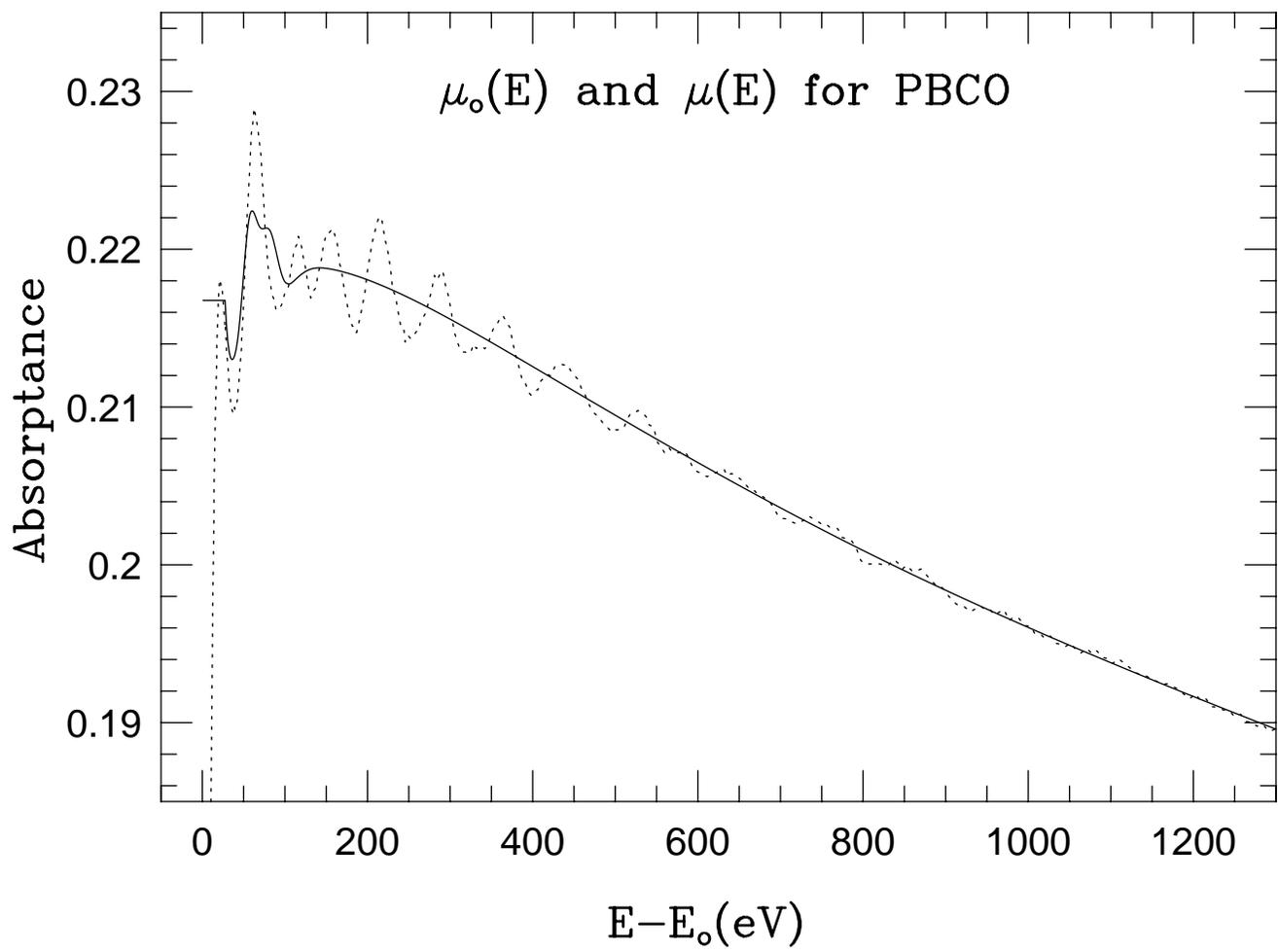
\end{document}